\documentclass[11pt]{article}
\usepackage{amssymb}
\usepackage{amsmath}
\usepackage{amscd}
\usepackage{latexsym}

\catcode`@=11
\renewcommand{\section}{\@startsection{section}{1}{0pt}{\medskipamount}
{\medskipamount}{\large\bf}}
\numberwithin{equation}{section}
\catcode`@=12

\newcommand{\Z}{\mathbb Z}
\newcommand{\C}{\mathbb C}
\newcommand{\R}{\mathbb R}
\newcommand{\cp}{{\mathbb C}P}
\newcommand{\Acal}{{\cal A}}
\newcommand{\Fcal}{{\cal F}}
\newcommand{\Ecal}{{\cal E}}

\newcommand{\Ncal}{{\cal N}}
\newcommand{\Ocal}{{\cal O}}
\newcommand{\Pcal}{{\cal P}}
\newcommand{\Jcal}{{\cal J}}
\newcommand{\Ucal}{{\cal U}}

\newcommand{\jfr}{{\mathfrak j}}
\newcommand{\Gfr}{{\mathfrak G}}

\newcommand{\al}{\alpha}
\newcommand{\be}{\beta}
\newcommand{\ga}{\gamma}
\newcommand{\de}{\delta}

\newcommand{\ve}{\varepsilon}

\newcommand{\la}{\lambda}
\newcommand{\na}{\nabla}

\newcommand{\we}{{\wedge}}
\def\im{\mathrm{i}}

\def\N={$\Ncal{=}$}
\def\pa{\mbox{$\partial$}}
\def\diff{\mathrm{d}}
\def\tr{\mathrm{tr}}
\def\sfrac#1#2{{\textstyle\frac{#1}{#2}}}
\def\>{\rangle}
\def\<{\langle}
\def\+{\dagger}
\def\={\ =\ }
\def\und{\quad\textrm{and}\quad}
\def\and{\qquad\textrm{and}\qquad}

\def\with{\quad\textrm{with}\quad}
\def\for{\quad\textrm{for}\quad}
\def\tU{\mathrm{U}}
\def\tSO{\mathrm{SO}}
\def\tSU{\mathrm{SU}}
\def\CS{\mathrm{CS}}
\def\Id{\mathrm{Id}}
\def\Re{\mathrm{Re}}
\def\Imm{\mathrm{Im}}
\def\gr{\mathrm{gr}}
\newcommand{\da}{\dot\alpha}
\newcommand{\db}{\dot\beta}

\begin{document}

\begin{titlepage}
\setcounter{page}{0}

\phantom{.}
\vskip 1.5cm

\begin{center}
{\LARGE{\bf
A Twistor Space Action for Yang-Mills Theory}}

\vspace{20mm}

{\Large Alexander D. Popov
}\\[10mm]

\noindent {\em Institut f\"ur Theoretische Physik,
Leibniz Universit\"at Hannover \\
Appelstra\ss{}e 2, 30167 Hannover, Germany
}\\
{Email: alexander.popov@itp.uni-hannover.de}

\vspace{15mm}

\begin{abstract}
\noindent We consider the twistor space $\Pcal^6\cong\R^4{\times}\cp^1$ of $\R^4$ with a non-integrable almost complex structure $\Jcal$ such that the canonical bundle of the almost complex manifold $(\Pcal^6, \Jcal )$ is trivial. It is shown that $\Jcal$-holomorphic Chern-Simons theory on a real $(6{\mid}2)$-dimensional graded extension $\Pcal^{6\mid 2}$ of the twistor space $\Pcal^6$ is equivalent to self-dual Yang-Mills theory on Euclidean space $\R^4$ with Lorentz invariant action. It is also shown that adding a local term to a Chern-Simons-type action on  $\Pcal^{6\mid 2}$, one can extend it to a twistor action describing full Yang-Mills theory. 
\end{abstract}

\vspace{12mm}


\end{center}
\end{titlepage}

\section{Introduction}

Let $M^4$ be an oriented real four-manifold with a Riemannian metric and $P(M^4, \tSO(4))$ the principal bundle of orthonormal frames over $M^4$. The {\it twistor space} Tw($M^4$) of $M^4$ can be defined as an associated bundle~\cite{AHS}
\begin{equation}\label{1.1}
{\rm Tw}(M^4) = {P}\times_{\tSO(4)}\tSO(4)/\tU(2)
\end{equation}
with the canonical projection
\begin{equation}\label{1.2}
\pi :\quad {\rm Tw}(M^4) \longrightarrow M^4\ .
\end{equation}
Fibres of this bundle are two-spheres $S^2_x\cong \tSO(4)/\tU(2)$ which parametrize complex structures $J_x$ on the tangent space $T_xM^4$ at $x\in M^4$ compatible with a Euclidean metric and orientation of $M^4$. It means that $J_x\in{\rm End}(T_xM^4)$ with $J_x^2=-\Id$ and $J_x$ is an isometry of $T_xM^4$ preserving orientation.

An almost complex structure $J$ on $M^4$  is a global section of the bundle  (\ref{1.2}). Note that while a manifold $M^4$ admits in general no almost complex structure (e.g. four-sphere $S^4$), its twistor space  Tw($M^4$) can always be equipped with two natural almost complex structures. The first, $\Jcal = \Jcal_+$, introduced in~\cite{AHS}, is {\it integrable} if and only if the Weyl tensor of Riemannian metric on  $M^4$ is self-dual, while the second, $\Jcal = \Jcal_-$, introduced in~\cite{ES}, is {\it non-integrable} (and never integrable), i.e. the Nijenhuis tensor of $\Jcal$ does not vanish.

Twistor space  $\Pcal^6=\,$Tw$(\R^4)\cong\R^4{\times}S^2$ of $\R^4$ with an almost complex structure $\Jcal$ is a particular case of almost complex six-manifolds to be discussed in this paper. Twistor space $(\Pcal^6, \Jcal )$ is a complex manifold $\Pcal^3_\C$ for integrable $\Jcal$ and it is an almost complex manifold with an SU(3)-structure and non-vanishing torsion for non-integrable $\Jcal$. Twistor literature focuses on complex twistor space $\Pcal^3_\C$ (see e.g.~\cite{PenRin, WardWells, MasWood}) and very rarely on the non-integrable case (see e.g.~\cite{ES, XuPhD, Popov1}).

The goal of twistor theory is to take some unconstraint analytic object on Tw($M^4$) (e.g. Dolbeault cohomology classes) and transform them to objects on $M^4$ which will be constrained by some differential equations~\cite{PenRin, WardWells}. In particular, the self-dual Yang-Mills (SDYM) equations on Euclidean space $\R^4$ can be described as field equations of holomorphic Chern-Simons theory defining holomorphic bundles on the complex twistor space  $\Pcal^3_\C$ via the Penrose-Ward correspondence~\cite{PenRin, WardWells, MasWood}. This correspondence can be extended to the non-integrable case (see e.g.~\cite{XuPhD, Popov1}).

The field equations of $\Jcal$-holomorphic Chern-Simons  ($\Jcal$-hCS) theory on $(\Pcal^6, \Jcal )$ read
\begin{equation}\label{1.3}
\Fcal^{0,2}=  P^{0,1} P^{0,1}\Fcal = (\diff\Acal + \Acal\wedge\Acal )^{0,2}=0   \ ,
\end{equation}
where $ P^{0,1}=\sfrac12 (\Id+\im\Jcal )$ is the projector onto (0,1)-part of one-forms, $\Acal$ is a connection one-form on a complex vector bundle $\Ecal$ over $(\Pcal^6, \Jcal )$ and $\Fcal =\diff\Acal + \Acal\wedge\Acal$ is the curvature of $\Acal$. One can expect that equations  (\ref{1.3}) are obtained by variation of the action functional 
\begin{equation}\label{1.4}
S=\frac{\im}{8}\int_{\Pcal^6}\Omega\wedge\CS(\Acal )^{0,3}=\frac{\im}{8}\int_{\Pcal^6}\Omega\wedge\tr(\Acal\wedge\diff\Acal + \sfrac{2}{3}\Acal\wedge\Acal\wedge\Acal )^{0,3}\ ,
\end{equation}
where $\Omega$ is a (3,0)-form w.r.t. $\Jcal$ on $(\Pcal^6, \Jcal )$, i.e. $\Omega$ is a global section of the canonical bundle of $(\Pcal^6, \Jcal )$. However, the canonical bundle of $\Pcal^3_\C\cong\C P^3\setminus \C P^1$ is the non-trivial holomorphic line bundle 
$\Ocal(-4)$ with the first Chern class -4. Hence, there is no non-singular holomorphic volume form $\Omega$ on $\Pcal^3_\C$. Thus, the functional  (\ref{1.4}) is not defined on $\Pcal^3_\C$.

The triviality of the canonical bundle can be restored if instead of $\Pcal^3_\C$ one considers the supertwistor space $\Pcal^{3\mid 4}_\C\cong\cp^{3\mid 4}\setminus\cp^{1\mid 4}$ with four holomorphic fermionic dimensions, each of type $\Pi\Ocal(1)$ bundle, where the operator $\Pi$ inverts the Grassmann parity of fibre coordinates. The canonical bundle of $\Pcal^{3\mid 4}_\C$ is trivial and hence there is a holomorphic volume form $\widetilde\Omega$ on $\Pcal^{3\mid 4}_\C$.  This fact was used by Witten for introducing twistor string theory and holomorphic Chern-Simons theory (hCS) on $\Pcal^{3\mid 4}_\C$~\cite{Witt}.
The action of hCS theory on $\Pcal^{3\mid 4}_\C$ can be written in the form  (\ref{1.4}) after substituting $\widetilde\Omega$ instead of $\Omega$ and integrating over $\Pcal^{3\mid 4}_\C$. The field equations will be (\ref{1.3}) with $\Acal^{0,1}=P^{0,1}\Acal$ depending on four Grassmann variables taking values in the bundle $\Pi\Ocal(1)\otimes\C^4$ over  $\Pcal^3_\C$. This hCS theory on $\Pcal^{3\mid 4}_\C$ in turn is equivalent~\cite{Witt} to self-dual subsector of \N=4 supersymmetric Yang-Mills theory on $\R^4$ (see e.g.~\cite{SaPhD, WolfPhD, Wolf1} for reviews and references) in the form of Chalmers and Siegel~\cite{ChSi}.  The \N=4 SDYM equations can be truncated to the bosonic SDYM equations~\cite{ChSi} and on the twistor level this was discussed e.g. in~\cite{LePo, Wolf2, Saem}. 

Despite the success of the supertwistor description of supersymmetric Yang-Mills theories, there was a desire to get a twistor description of pure bosonic SDYM theory. Recently, it was proposed by Costello to work with hCS theory on the bosonic twistor space  $\Pcal^3_\C$ by allowing $\Omega$ in  (\ref{1.4}) to be meromorphic instead of holomorphic~\cite{Costello}. After choosing a meromorphic form $\Omega$ on  $\Pcal^3_\C$ and imposing some boundary conditions on fields at poles of $\Omega$, one can reduce the action  (\ref{1.4}) to the $4d$ action for SDYM theory as it was demonstrated in~\cite{Costello, Skinner}.  Depending on the gauge choice, the twistor action is reduced to the action for group-valued fields~\cite{Don, NS} or to the action for Lie-algebra valued fields~\cite{LeMu, Parkes}, both of which are well known in the literature. However, the choice of (3,0)-form $\Omega$ and of its singularities is not unique and different choices lead to a range of actions on $\R^4$, not all of which have equations of motion equivalent to the SDYM equations~\cite{Skinner}. 

All the above-mentioned actions break Lorentz invariance. The actions~\cite{Don}-\cite{Parkes} for the SDYM equations were discussed long time ago by Chalmers and Siegel in~\cite{ChSi}, where it was shown that these $4d$ actions at more than one loop generate diagrams that do not relate to quantum Yang-Mills theory. These flaws are absent for the Chalmers-Siegel $4d$ action which is a truncation (a limit of small coupling constant) of the standard Yang-Mills action. We want to obtain this $4d$ action in the framework of twistor approach. We show that this is possible by using a non-integrable almost complex structure $\Jcal$ on the twistor space $\Pcal^6$ such that the canonical bundle becomes trivial and hence there exists a globally defined (3,0)-form $\Omega$ on $(\Pcal^6, \Jcal)$ which can be used in  (\ref{1.4}).

The action~\cite{ChSi} contains gauge field coupled with a propagating anti-self-dual auxiliary field $G_{\da\db}=\ve^{\al\be} G_{\al\da , \be\db}$ with $\da , \db =1,2$. The field $G_{\da\db}$ corresponds to additional degrees of freedom parametrized by some cohomology groups on the complex twistor space $\Pcal^3_\C$~\cite{Wood, Wolf1} and can be obtained from the component $\Acal^{0,1}$ along $\C P^1\hookrightarrow \Pcal^{3\mid 4}_\C$ in hCS theory on the supertwistor space (see e.g.~\cite{Wolf1} and references therein). This $G_{\da\db}$ enters into the \N=4 SDYM supermultiplet $(f_{\al\be}, \chi^{\al i}, \phi^{ij}, \tilde\chi_{\da i}, G_{\da\db})$, where the fields have helicities $(+1, +\sfrac12, 0, -\sfrac12, -1)$, $i=1,...,4$. Truncations of the self-dual \N=4 super-Yang-Mills to the case $\Ncal <4$, including the bosonic case \N=0, can be obtained by considering weighted projective supertwistor space~\cite{Wolf2, WolfPhD} or exotic supertwistor space~\cite{Saem, SaPhD}. The approach similar to that in~\cite{Wolf2, Saem} can be used in the case of twistor space $(\Pcal^6, \Jcal)$ with non-integrable almost complex structure $\Jcal$ on $\Pcal^6$. We will show that the $4d$ Chalmers-Siegel action~\cite{ChSi} can be obtained from an action functional for $\Jcal$-hCS theory on a graded twistor space $\Pcal^{6\mid 2}$ with two real fermionic directions, each parametrizing trivial real line bundle over $(\Pcal^6, \Jcal)$. The Chern-Simons type action on $\Pcal^{6\mid 2}$ is introduced by using globally defined form $\widetilde\Omega = \Omega\we\diff\eta_1\we\diff\eta_2$ on $\Pcal^{6\mid 2}$, where $\Omega$ is a global section of the trivial canonical bundle of $\Pcal^{6}$. Components of gauge potential $\Acal$ in this theory take values in the Grassmann algebra $\Lambda (\R^2)$ generated by two real scalars $\eta_1, \eta_2$. We also show that this action can be extended to a twistor action describing full Yang-Mills theory on $\R^4$ after adding some local terms to $\Jcal$-hCS Lagrangian on the twistor space $\Pcal^{6\mid 2}$.

\section{Self-dual Yang-Mills and twistors}

\noindent {\bf Almost complex structures on Tw($M^4$).} We defined the twistor space  Tw($M^4$) of a Riemannian manifold $M^4$ as the associated bundle  (\ref{1.1}) of complex structures $J_x$ on tangent spaces $T_xM^4$. Global sections of the projection (\ref{1.2}) are identified, if such sections exist, with almost complex structures $J$ on $M^4$, i.e. with tensors $J=(J^\nu_\mu)\in\,$End$(TM^4)$ such that $J^\sigma_\mu J^\nu_\sigma=-\de^\nu_\mu ,
\mu , \nu =1,...,4.$

While a manifold $M^4$ has in general no almost complex structures, its twistor space $Q^6:=$Tw$(M^4)$ can be always provided in a natural way with an almost complex structure $\Jcal$, a tensor on $Q^6$ with $\Jcal^2=-\Id$. In fact, the Levi-Civita connection on $M^4$ generates the splitting of the tangent bundle $TQ^6$ into the direct sum
\begin{equation}\label{2.1} 
TQ^6=V\oplus H
\end{equation}
of vertical and horizontal subbundles of  $TQ^6$. The space $V_q$ in $q\in Q^6$ is tangent to the fibre $\pi^{-1}(\pi (q))$ over $x=\pi (q)\in M^4$ of the projection $\pi: Q^6\to M^4$. Recall that the fibre over $x=\pi (q)$ is identified with $S^2_x\cong \tSO(4)/\tU(2)$ and so it has a natural complex structure $J^v$. Hence, we can define an almost complex structure $\Jcal$ on $Q^6$ using the decomposition  (\ref{2.1}) by setting 
\begin{equation}\label{2.2} 
\Jcal=\Jcal^{\rm{\rm{int}}}=\Jcal^{v}\oplus \Jcal^{h}\  ,
\end{equation}
where $\Jcal^{h}$ is an almost complex structure equal in the point $q\in Q^6$ to the complex structure $\Jcal^{h}_q$ on $H_q\cong T_{\pi (q)}M^4=T_{x}M^4$. Thus, the twistor space $Q^6$ has a natural almost complex structure $\Jcal$. 

It was shown in~\cite{AHS} that if the Weyl tensor of $M^4$ is self-dual then the almost complex structure  (\ref{2.2}) on  $Q^6$ is integrable and $(Q^6, \Jcal^{\rm{int}})$ inherits the structure of a complex analytic 3-manifold $Q^3_\C$. It was also shown in~\cite{ES} that 
\begin{equation}\label{2.3} 
\Jcal=\Jcal^{\rm{\rm{non}}}=\Jcal^{v}\oplus(- \Jcal^{h})
\end{equation}
is an almost complex structure on $Q^6$ which is never inegrable. These structures differ in sign along $M^4$. 

\medskip

\noindent {\bf Twistor corespondence.} Let $E$ be a rank $k$ complex vector bundle over $M^4$ and $A$ a connection one-form (gauge potential) on $E$ with the curvature $F=\diff A + A\wedge A$ (gauge field). The gauge field $F$ is called self-dual if it satisfies the equations
\begin{equation}\label{2.4} 
\ast F=F \quad\Leftrightarrow\quad\sfrac12\, \ve_{\mu\nu\lambda\sigma}F^{\lambda\sigma}=F_{\mu\nu}\  ,
\end{equation}
where $\ast$ denotes the Hodge star operator, $\ve_{\mu\nu\lambda\sigma}$ is the completely antisymmetric tensor on $M^4$ with $\ve_{1234}=1$ in the Riemannian metric $\diff s^2=\delta_{\mu\nu}e^\mu e^\nu$ for an orthonormal basis $\{e^\mu\}$ on $T^*M^4$.

Bundles $E$ with self-dual connections $A$ are called self-dual. It was proven in~\cite{AHS} that the self-dual bundle $E$ over self-dual manifold $M^4$ lifts to a holomorphic bundle $\Ecal$ over the complex twistor space $Q^3_\C =($Tw$(M^4), \Jcal^{\rm{\rm{int}}})$ and $\Ecal$ is holomorphically trivial on fibres $\C P^1_x$ of projections $\pi : Q^6\to M^4$ for each $x\in M^4$. The bundle $\Ecal = \pi^*E$ is defined by the connection $\Acal = \pi^*A$ such that its curvature $\Fcal =\diff\Acal + \Acal\we \Acal$ satisfies the equations (\ref{1.3}) and $\Fcal =\pi^* F$ is the pull-back to $\Ecal$ of self-dual gauge field $F$ on $E\to M^4$. Vice versa, solutions to the holomorphic Chern-Simons field equations  (\ref{1.3})  on the twistor space $Q^3_\C$, with $\Fcal^{}_{|\C P^1_x}=0$ for any $x\in M^4$, give solutions to the SDYM equations   (\ref{2.4})  on $M^4$. The map between solutions to the SDYM equations on $M^4$ and solutions to the hCS field equations on $Q^3_\C =($Tw$(M^4), \Jcal^{\rm{int}})$ is called the Penrose-Ward transform.

For non-integrable almost complex structure   (\ref{2.3})  on $Q^6$ the manifold $(Q^6, \Jcal^{\rm{\rm{non}}})$ is not complex. However, on  $(Q^6, \Jcal^{\rm{\rm{non}}})$ one can introduce bundles with $\Jcal$-holomorphic structure (pseudo-holomorphic bundles)~\cite{Bryant}. Let $\Ecal$ be a complex rank $k$ vector bundle over $Q^6$ endowed with a connection  $\Acal$. According to Bryant~\cite{Bryant}, a connection  $\Acal$ on $\Ecal$ is said to define a $\Jcal$-holomorphic structure if it has curvature $\Fcal$ of type (1,1) w.r.t. $\Jcal$, i.e.
\begin{equation}\label{2.5} 
\Fcal^{0,2}=0\  .
\end{equation}
It is not difficult to show that twistor correspondence between solutions of SDYM equations  (\ref{2.4}) on $M^4$ and solutions of  $\Jcal$-hCS equations  (\ref{2.5}) on the almost complex twistor space $(Q^6, \Jcal)$ still persists (see e.g.~\cite{Popov1}). This will be discussed in more details later for the case of flat Euclidean space $M^4=\R^4$.

\section{Twistor space of $\R^4$}

According to the definition (\ref{1.1}), twistor space of $\R^4$ is $\Pcal^6{:=}\,$Tw$(\R^4)\cong\R^4\times S^2$. Due to diffeomorphism with $\R^4\times S^2$, the manifold $\Pcal^6$ is fibred not only over $\R^4$,
\begin{equation}\label{3.1}
\pi:\quad \Pcal^6\stackrel{S^2}{\longrightarrow} \R^4\ ,
\end{equation}
but also over $S^2$,
\begin{equation}\label{3.2}
\Pcal^6\stackrel{\R^4}{\longrightarrow} S^2\ ,
\end{equation}
with spaces $\R^4$ as fibres.

\medskip

\noindent {\bf Almost complex structures $\Jcal$.} In section 2 we described generic construction of an almost complex structure $\Jcal$ on a twistor space Tw$(M^4)$. Here, we give explicit form of $\Jcal$ for the case $M^4=\R^4$. 

Recall that a complex structure $J$ on $\R^4$ is a tensor $J=(J^\nu_\mu )$ such that $J^\sigma_\mu J^\nu_\sigma =-\de^\nu_\mu$. All constant complex structures on $\R^4$ are parametrized by the two-sphere $S^2\cong \tSO(4)/\tU(2)\cong \tSU(2)/\tU(1)$ defined by the equation 
\begin{equation}\label{3.3}
\de_{ab}s^as^b =1
\end{equation}
for $s^a\in\R^3,\ a,b=1,2,3$. One can choose generic $J$ in the form 
\begin{equation}\label{3.4}
J^\nu_\mu = s_a\bar\eta^a_{\mu\sigma}\de^{\sigma\nu}\ ,
\end{equation}
where
\begin{equation}\label{3.5}
\bar\eta^a_{\mu\nu}=\left\{\ve^a_{bc},\ \mu =b, \nu =c; \quad -\de^a_\mu,\ \nu =4;\quad \de^a_\nu,\ \mu =4\right\}
\end{equation}
are antisymmetric 't~Hooft tensors, $\mu , \nu =1,...,4$. Using the identities
\begin{equation}\label{3.6}
\bar\eta^a_{\mu\sigma}\bar\eta^b_{\sigma\nu} = -\de^{ab}\de_{\mu\nu}-\ve^{abc}\bar\eta^c_{\mu\nu}\ ,
\end{equation}
one can show that $J^2=-\Id$. Here, we consider $\R^4$ as a space with the metric $\diff s^2_{\R^4}=\de_{\mu\nu}\diff x^\mu \diff x^\nu$, where $x^\mu$ are coordinates on $\R^4$.

Let $\{e^\al\}$ represents an orthonormal coframe on $S^2$, i.e.
\begin{equation}\label{3.7}
\diff s^2_{S^2}=\de_{\al\be}e^\al e^\be
\end{equation}
for $\al , \be =1,2$. The canonical form of complex structure $\jfr$ on $S^2$ is
\begin{equation}\label{3.8}
\jfr = (\jfr^\be_\al )\with \jfr^2_1=-\jfr^1_2=1\quad\Rightarrow\quad\jfr^\sigma_\al\jfr_\sigma^\be =-\de^\be_\al\ .
\end{equation}
It is obvious that both
\begin{equation}\label{3.9}
 \Jcal = \Jcal^{\rm{int}}=(J, \jfr )
\end{equation}
and
\begin{equation}\label{3.10}
 \Jcal = \Jcal^{\rm{non}}=(-J, \jfr )
\end{equation}
are almost complex structures on the twistor space $\Pcal^6$ of $\R^4$. Complex twistor space $\Pcal^3_\C =(\Pcal^6, \Jcal )$ with integrable almost complex structure $\Jcal =\Jcal^{\rm{int}}$ has been studied a lot in the literature and in the following we will focuse on non-integrable almost complex structure $ \Jcal = \Jcal^{\rm{non}}$.

\medskip

\noindent {\bf Complex coordinates for  $\Jcal =\Jcal^{\rm{int}}$.} The two-sphere $S^2$, global coordinates $s^a$ on which are used in (\ref{3.4}), is conformally equivalent to $\R^2$. One can cover $S^2$ by two patches $U_{\pm}\cong \R^2$ with local coordinates 
\begin{equation}\label{3.11}
v_+^\al =\frac{s^\al}{1+s^3}\ \ {\mbox {on}}\ \ U_+\and v_-^\al =\frac{s^\al}{1-s^3}\ \ {\mbox {on}}\ \ U_-\ ,
\end{equation}
in which the metric on $S^2$ is conformally flat,
\begin{equation}\label{3.12}
\diff s^2_{S^2|U_\pm}=\de_{\al\be} e^\al_\pm e^\be_\pm = \frac{4\de_{\al\be}\diff v^\al_\pm\diff v^\be_\pm}{(1+\rho^2_\pm)^2}\with
\rho^2_{\pm}=\de_{\al\be} v^\al_{\pm} v^\be_{\pm} \ .
\end{equation}
 On the intersection of two patches we have
\begin{equation}\label{3.13}
v^\al_+=\rho^2_+v^\al_-\ ,
\end{equation}
where $\al , \be =1,2$.

On $S^2$ one can introduce vector fields of type (1,0) and (0,1) w.r.t. $\jfr$ from (\ref{3.8}), 
\begin{equation}\label{3.14}
\frac{\partial}{\partial\la_{\pm}}\and \frac{\partial}{\partial\bar\la_{\pm}}\ ,\qquad \jfr(\partial^{}_{\la_{\pm}})=\im \partial^{}_{\la_{\pm}}\und\jfr(\partial^{}_{\bar\la_{\pm}})=-\im \partial^{}_{\bar\la_{\pm}}\ ,
\end{equation}
where 
\begin{equation}\label{3.15}
\la_{\pm}=v^1_{\pm} +\im v^2_{\pm}\and \la_+=\la_-^{-1}\ \ {\mbox {on}}\ \  U_+\cap U_-
\end{equation}
are complex coordinates on $U_\pm\subset S^2$. One-forms, dual to the vector fields (\ref{3.14}), are $\diff\la_\pm$ and  $\diff\bar\la_\pm$.
Sphere $(S^2, \jfr )$ with the coordinates  (\ref{3.15}) can be identified with the Riemann sphere $\cp^1$.

By using the complex structure  (\ref{3.4}) on $\R^4$, one can introduce a $\cp^1$-family of complex coordinates on $\R^4$ given by formulae
\begin{equation}\label{3.16}
w^1_+=y^1 + \la_+\bar y^2\and w^2_+=y^2 - \la_+\bar y^1\ ,
\end{equation}
where
\begin{equation}\nonumber
y^1=x^1+\im x^2\ ,\ \  y^2=x^3-\im x^4\ ,\ \ \bar y^1=x^1-\im x^2\und  \bar  y^2=x^3+\im x^4\ .
\end{equation}
The coordinates  (\ref{3.15}) together with  (\ref{3.16}) provide complex coordinates on $\Pcal^6$ given by
\begin{equation}\label{3.17}
w^1_+ \ , \ \ w^2_+\und w^3_+=\la_+\quad \mbox{on}\quad\Ucal_+=U_+\times\R^4\subset\Pcal^6 
\end{equation}
and
\begin{equation}\nonumber
w^1_-=\la_-y^1 + \bar y^2\ ,\  \  w^2_-=\la_- y^2 - \bar y^1\und w^3_-=\la_-\ \ {\mbox {on}}\ \  \Ucal_-=U_-\times\R^4\subset\Pcal^6\ .
\end{equation}
On the intersection $\Ucal_+\cap\Ucal_-$ of patches $\Ucal_\pm\subset\Pcal^6$ these coordinates are related by formulae
\begin{equation}\label{3.18}
w_+^\al =w_+^3w_-^\al\and w_+^3=\frac{1}{w_-^3}\ \ {\mbox {on}}\ \  \Ucal_+\cap\Ucal_-\ .
\end{equation}
Hence, the transition functions relating $w_+^a$ and $w_-^a$ are holomorphic functions on $\Ucal_+\cap\,\Ucal_-,\  a=1,2,3.$ This means that $\Jcal^{\rm{int}}$ is an integrable almost complex structure and $\Pcal_\C^3=(\Pcal^6, \Jcal^{\rm{int}})$ is a complex 3-manifold. From  (\ref{3.16}) - (\ref{3.18})  it follows that the manifold  $\Pcal_\C^3$ can be identified with the total space of the holomorphic vector bundle over $\cp^1$,
\begin{equation}\label{3.19}
\Pcal_\C^3=\Ocal (1)\oplus\Ocal(1)\quad\longrightarrow\quad\cp^1\ ,
\end{equation}
with coordinates $w^\al_\pm$ on fibres $\C^2_J$ over points $J\in\cp^1$ parametrized by $\la_\pm\subset U_\pm\subset\cp^1$.

\medskip

\noindent {\bf Complex coordinates for  $\Jcal =\Jcal^{\rm{non}}$.} By using the almost complex structure  (\ref{3.10}), we can introduce complex coordinates 
\begin{equation}\label{3.20}
z^1_+=\bar w^1_+=\bar y^1 + \bar\la_+y^2\ ,\ \  z^2_+=\bar w^2_+=\bar y^2 - \bar\la_+y^1\ ,\ \ z^3_+= w^3_+=\la_+\ \ {\mbox {on}}\ \  \Ucal_+\subset\Pcal^6 
\end{equation}
and
\begin{equation}\label{3.21}
z^1_-=\bar w^1_-=\bar\la_-\bar y^1 + y^2\ ,\ \  z^2_-=\bar w^2_-=\bar\la_-\bar y^2 - y^1\ ,\ \ z^3_-= w^3_-=\la_-\ \ {\mbox {on}}\ \  \Ucal_-\subset\Pcal^6 \ .
\end{equation}
On the intersection $\Ucal_+\cap\Ucal_-$ of two coordinate patches $\Ucal_\pm\subset\Pcal^6=\Ucal_+\cup\Ucal_-$ we have
\begin{equation}\label{3.22}
z^\al_+=\bar z^3_+z^\al_-\and z^3_+=\frac{1}{z^3_-}\ .
\end{equation}
From (\ref{3.22}) we see that the transition functions on $\Ucal_+\cap\Ucal_-$ are not holomorphic. This reflects non-integrability of the almost complex structure   (\ref{3.10}). From   (\ref{3.22}) it follows that the manifold $(\Pcal^6, \Jcal)$ with  $\Jcal =\Jcal^{\rm{non}}$ can be identified with the total space of the anti-holomorphic vector bundle
\begin{equation}\label{3.23}
\bar\Ocal (1)\oplus\bar\Ocal (1)\quad\longrightarrow\quad\cp^1
\end{equation}
over $\cp^1$. Both base and fibres $\bar\C^2_J$ of this bundle are complex spaces but they do not glue into a complex manifold for $\Jcal$ given by (\ref{3.10}).

\medskip

\noindent {\bf Spinor notation}. The rotation group SO(4) of space $\R^4$ is locally isomorphic to the group SU(2)$\times$SU(2), where both groups SU(2) have two-dimensional fundamental (spinor) representations
\begin{equation}\label{3.24}
\mu =(\mu_\al )\and \la =(\la_{\dot\al})\ .
\end{equation}
Commuting components $\la_{\da}$ of the spinor $\la$ are homogeneous coordinates on the Riemannian sphere $\cp^1$ such that
\begin{equation}\label{3.25}
\frac{\la_{\dot 2}}{\la_{\dot 1}}=:\la_+\ \ {\mbox {on}}\ \  U_+\subset \cp^1\and \frac{\la_{\dot 1}}{\la_{\dot 2}}=:\la_-\ \ {\mbox {on}}\ \  U_-\subset \cp^1\ .
\end{equation}
Obviously, $\la_+=\la_-^{-1}$ if $\la_{\dot 1}\ne 0$ and $\la_{\dot 2}\ne 0$. 

Isomorphism SO(4)$\cong$SU(2)$\times$SU(2) allows also to introduce spinor notation for complex coordinates on $\R^4$ by formula 
\begin{align}\label{3.26}
&\hspace{3cm}(x^{\al\da})=\left(\begin{matrix}x^{1\dot{1}}&x^{1\dot{2}}\\
x^{2\dot{1}}&x^{2\dot{2}}\end{matrix}\right)=
\left(\begin{matrix}y^{1}&-\bar y^{2}\\
y^{2}&\bar y^{1}\end{matrix}\right)=
\left(\begin{matrix}x^{1}+\im x^2&-(x^3+\im x^4)\\
x^3-\im x^4&x^{1}-\im x^2\end{matrix}\right)\ .
\end{align}
From   (\ref{3.26}) it follows that
\begin{equation}\label{3.27}
x^{1\dot{1}}=\bar x^{2\dot{2}}\and x^{1\dot{2}}=-\bar x^{2\dot{1}}\ ,
\end{equation}
where the overbar denotes complex conjugation. By using   (\ref{3.26}), one can rewrite   (\ref{3.16})  and (\ref{3.20}) as follows
\begin{equation}\label{3.28}
w_+^{\al} = x^{\al\da}\la^+_{\da}\and z_+^{\al} =-\jfr^{\al}_{\be}\, x^{\be\db}\,\hat{\la}^+_{\db}\ ,
\end{equation}
where
\begin{equation}\label{3.29}
(\la_{\da}^+)=\frac{1}{\la_{\dot{1}}}\, (\la_{\da})=\begin{pmatrix} 1\\ \la_+ \end{pmatrix}  \and 
(\hat\la_{\da}^+)=\begin{pmatrix} 0&-1\\ 1&0 \end{pmatrix} (\bar \la_{\da}^+)=\begin{pmatrix}-\bar \la_+\\ 1 \end{pmatrix}\ .
\end{equation}
By definition, we have $\la_{\da}^-=\la_+^{-1}\la_{\da}^+$. and  $\hat \la_{\da}^-=\bar\la_+^{-1}\hat\la_{\da}^+$.

\medskip

\noindent {\bf Vector fields and one-forms}. On the twistor space $(\Pcal^6, \Jcal )$ with $\Jcal$ from  (\ref{3.10}) we have the natural basis
$\left\{\frac{\pa}{\pa z^a_\pm}\right\}$ for the space of (1,0) vector fields. On the intersection we have
\begin{equation}\label{3.30}
\frac{\pa}{\pa z^\al_+} = \bar z_-^3\frac{\pa}{\pa z^\al_-}\and \frac{\pa}{\pa z^3_+} = -(z_-^3)^2\frac{\pa}{\pa z^3_-}-
z_-^3\bar  z^\al_-\frac{\pa}{\pa \bar z^\al_-}\ .
\end{equation} 
Using formulae  (\ref{3.28}), we can express these vector fields in terms of coordinates $(x^{\al\dot{1}}, \la_\pm )$ and their complex conjugates according to
\begin{eqnarray}\nonumber
\frac{\pa}{\pa z^\al_\pm}& = &-\ga_{\pm}\ \jfr_{\al}^{\be}\ \la^{\db}_{\pm}\ \frac{\pa}{\pa x^{\be\db}}=:-\jfr_{\al}^{\be}\ V^{\pm}_{\be}\ ,\\\label{3.31}
\frac{\pa}{\pa z^3_+} &=&\frac{\pa}{\pa\la_+} +\ga_+\,\jfr_{\al}^{\be}x^{\al\dot{1}}\,V^+_\be \ ,\quad
\frac{\pa}{\pa z^3_-} =\frac{\pa}{\pa\la_-} -\ga_-\,\jfr_{\al}^{\be}\,x^{\al\dot{2}}V^-_{\db} \ ,
\end{eqnarray}
where we have used
\begin{equation}\label{3.32}
\la_\pm^{\da}=\ve^{\da\db}\la^\pm_{\db}\with \ve^{\dot{1}\dot{2}}=-\ve^{\dot{2}\dot{1}}=1\und \ga_\pm=\frac{1}{1+\la_\pm\bar\la_\pm}=\frac{1}{\hat\la_\pm^{\da}\la^\pm_{\da}}
\end{equation} 
together with the convention $\ve_{\dot{1}\dot{2}}=-\ve_{\dot{2}\dot{1}}=-1$, which implies $\ve_{\da\db} \ve^{\dot{\be}\dot{\ga}}=\de^{\dot{\ga}}_{\da}$. Thus, the vector fields
\begin{equation}\label{3.33}
V^\pm_\al = \ga_\pm\, \la^{\da}_\pm\pa_{\al\da}\ ,\quad  V_3^+=\ga_+^{-2}\pa_{\la_+}\und V_3^-=\bar\la_+^{-2}V_3^+
\end{equation} 
can be chosen as a basis of vector fields of type (1,0) on $\Ucal_\pm\subset\Pcal^6$ in the coordinates $(x^{\al\da}, \la_\pm , \bar\la_\pm )$. Complex conjugate of (\ref{3.33}) provide us with the vector fields
\begin{equation}\label{3.34}
\bar V^\pm_\al = \ga_\pm\, \jfr^{\be}_\al\, \hat\la_\pm^{\db}\,\pa_{\be\db}\ ,\quad
\bar V_3^+=\ga_+^{-2}\pa_{\bar\la_+} \und \bar V_3^-=\la_+^{-2}\bar V_3^+
\end{equation} 
which form a basis of vector fields of type (0,1) on $\Ucal_\pm\subset\Pcal^6$.

It is easy to check that the basis of (1,0)- and (0,1)-forms on $\Ucal_\pm$, which are dual to the vector fields (\ref{3.33}) and (\ref{3.34}), is given by forms
\begin{eqnarray}\nonumber
E_\pm^\al &=&-(\diff x^{\al\da})\hat\la^\pm_{\da}\ , \quad E_+^3=\ga_+^2\diff\la_+\und E^3_-=\bar\la_+^{-2}E_+^3\ ,
\\ \label{3.35}
\bar E_\pm^\al &=&-\jfr_{\be}^\al\, (\diff x^{\be\db})\la_{\db}\ , \quad \bar E_+^3=\ga_+^2\diff\bar\la_+\und\bar  E^3_-=\la_+^{-2}\bar E_+^3\ .
\end{eqnarray}
One can easily verify that
\begin{equation}\label{3.36}
\diff_{|\Ucal_\pm}=\diff z_\pm^a\frac{\pa}{\pa z_\pm^a} + \diff\bar z_\pm^a\frac{\pa}{\pa\bar z_\pm^a} = E_\pm^a\,V^\pm_a + \bar E_\pm^a\,\bar V^\pm_a\ .
\end{equation} 

\medskip

\noindent {\bf Geometry of $(\Pcal^6, \Jcal )$}. 
We consider the twistor space $(\Pcal^6, \Jcal )$ with $\Jcal$ from (\ref{3.10}) and coordinates $\{z_\pm^a\}$ on $\Ucal_\pm\subset\Pcal^6$ given by  (\ref{3.20})-(\ref{3.22}). In the following we often omit the signs $\pm$ in coordinates, vector fields, one-forms etc. by considering all formulae on the patch $\Ucal_+\subset\Pcal^6$.

By direct calculations we obtain that nonzero commutators of vector fields  (\ref{3.33}) and (\ref{3.34}) are 
\begin{equation}\label{3.37}
[V_3, V_\al ] = -\ga^{-1}\,\jfr^\be_\al \bar V_\be \ ,\quad [V_3, \bar V_\al ] = -\bar\la\,\ga^{-1}\, \bar V_\al\ ,\quad
[V_3, \bar V_3] = 2\ga (\bar\la\,\bar V_3-\la\,V_3)\ ,
\end{equation} 
\begin{equation}\label{3.38}
[\bar V_3, \bar  V_\al ] = -\ga^{-1}\,\jfr^\be_\al V_\be 
 \und [\bar V_3, V_\al ] = -\la\ga^{-1}\,V_\al \ ,
\end{equation} 
where we used the formulae
\begin{equation}\label{3.39}
\pa_\la (\ga\la^{\da})=\ga^2\,\hat\la^{\da}\and \pa_{\bar\la} (\ga\hat\la^{\da})=-\ga^2\,\la^{\da}\ .
\end{equation} 
To prove integrability of an almost complex structure $\Jcal$ one has to show that commutators of vector fields of type (0,1) w.r.t. $\Jcal$ will again be  vector fields of type (0,1)~\cite{KobNom}. From (\ref{3.37}) we see that this is not the case and therefore $\Jcal$ is not integrable. For one-forms (\ref{3.35}) we have
\begin{eqnarray}\nonumber
\diff E^1 &=& \la\ga^{-1}\, \bar E^3\wedge E^1 + \ga^{-1}\, \bar E^2\wedge \bar E^3\ , \\\label{3.40}
\diff E^2 &=& \la\ga^{-1}\, \bar E^3\wedge E^2 + \ga^{-1}\, \bar E^3\wedge \bar E^1\ , \\\nonumber
\diff E^3 &=&-2 \la\ga^{-1}\, \bar E^3\wedge E^3\ , 
\end{eqnarray} 
and complex conjugate formulae. The first terms in  (\ref{3.40}) define a torsionful connection on $\Pcal^6$ with values in $u(1)\subset su(3)$ and the last terms define the Nijenhuis tensor (torsion) with non-vanishing components
\begin{equation}\label{3.41}
N^1_{\bar 2\bar 3}=\ga^{-1}\ ,\quad N^2_{\bar 3\bar 1}=\ga^{-1}
\end{equation} 
plus their complex conjugate $N^{\bar 1}_{23}=N^{\bar 2}_{31}=\ga^{-1}$. From  (\ref{3.40}) we again see that $(\Pcal^6, \Jcal )$ is not a complex manifold but the total space   (\ref{3.23}) of the anti-holomorphic bundle over $\cp^1$. Furthermore, from  (\ref{3.40}) we see that $(\Pcal^6, \Jcal )$ has an SU(3)-structure and the globally defined (3,0)-form $\Omega$ with
\begin{equation}\label{3.42}
\Omega = E^1_+\wedge  E^2_+ \wedge E^3_+ = E^1_-\wedge  E^2_- \wedge E^3_- \quad\mbox{on}\quad\Ucal_+\cap\Ucal_-
\end{equation} 
since 
\begin{equation}\label{3.43}
E^\al_+=\bar\la_+  E^\al_- \and E^3_+=\bar\la_+^{-2}  E^3_- \ .
\end{equation} 
Hence, the canonical bundle of  $(\Pcal^6, \Jcal )$ is trivial. From  (\ref{3.40}) it follows that
\begin{equation}\label{3.44}
\diff (\Imm\Omega )=0\ ,
\end{equation} 
\begin{equation}\label{3.45}
\diff (\Re\Omega )=-\ga^{-1}(E^1\wedge \bar E^1 + E^2\wedge \bar E^2)\wedge E^3\wedge \bar E^3\ ,
\end{equation} 
i.e. the real part of $\Omega$ is not closed. For the volume form on $\Pcal^6$ we have 
\begin{equation}\label{3.46}
\mbox{Vol}_6=\frac{\im}{8}\,\Omega\wedge\bar\Omega =-\frac{\im}{2}\,\diff^4x\wedge\frac{\diff\la\wedge\diff\bar\la}{(1+\la\bar\la)^2}\ ,
\end{equation} 
where $\diff^4x=\diff x^1\wedge\diff x^2\wedge\diff x^3\wedge\diff x^4$ in the coordinates (\ref{3.26}).

\medskip

\noindent {\bf Twistor correspondence}. To conclude this section we describe a twistor correspondence between the SDYM model on $\R^4$ and $\Jcal$-hCS theory on $(\Pcal^6, \Jcal )$.

Consider a complex vector bundle $E$ over $\R^4$ with a connection $A=A_{\al\da}\diff x^{\al\da}$  and the covariant derivative $\nabla =\diff x^{\al\da}(\pa_{\al\da} + A_{\al\da})$. Using the projection $\pi : \ \Pcal^6\to\R^4$ from  (\ref{3.1}), we can pull back the bundle $E$ to a bundle $\Ecal =\pi^*E$ with the pulled back connection $\Acal =\pi^*A$ and the covariant derivative $\widetilde\nabla =\pi^*\nabla$, whose (0,1)-component is\footnote{We are working on the patch $\Ucal_+=U_+\times\R^4\subset\Pcal^6$ and omit subscript \& superscript ``+''  in formulae.}
\begin{equation}\label{3.47}
\widetilde\nabla^{0,1}=\bar E^\al (\bar V_\al + \ga\,\jfr^\be_\al\,\hat\la^{\db}\, A_{\be\db}) + \bar E^3\bar V_3\ .
\end{equation} 
Equations  (\ref{2.5}) of $\Jcal$-holomorphic Chern-Simons theory on $(\Pcal^6, \Jcal )$ read
\begin{equation}\label{3.48}
[\widetilde\nabla^{0,1}_a,\ \widetilde\nabla^{0,1}_b] - \widetilde\nabla^{0,1}_{[\bar V_a, \bar V_b]}=0\ ,
\end{equation}
where $a=(\al , 3)=1,2,3$. Substituting   (\ref{3.47}) into   (\ref{3.48}) with
\begin{equation}\label{3.49}
\bar\Acal_\al = \ga\, \jfr^\be_\al\, \hat\la^{\db}\, A_{\be\db}\ ,\quad \bar\Acal_3 = 0\ ,\quad \Acal_\al = \ga\, \la^{\da}A_{\al\da}\und A_3=0\ ,
\end{equation}
we see that  (\ref{3.48}) are equivalent to the equations
\begin{equation}\label{3.50}
 \hat\la^{\da}\,  \hat\la^{\db}\,\left[\pa_{\al\da}+A_{\al\da}\ , \ \pa_{\be\db}+A_{\be\db}\right]= \hat\la^{\da}\,  \hat\la^{\db}\, F_{\al\da , \be\db}=0\ ,
\end{equation}
where $F=\diff A + A\wedge A$ is the curvature of $A$. Recall that in the spinor notation $F$ has the components
\begin{equation}\label{3.51}
F_{\al\da , \be\db}=\ve_{\da\db}f_{\al\be} + \ve_{\al\be}f_{\da\db}\ ,
\end{equation}
where symmetric tensors 
\begin{equation}\label{3.52}
f_{\al\be}=\sfrac12\,\ve^{\da\db}F_{\al\da , \be\db}\and  f_{\da\db}=\sfrac12\,\ve^{\al\be}F_{\al\da , \be\db}
\end{equation}
represent self-dual $F^+$ and anti-self-dual $F^-$ parts of the curvature $F=F^+ + F^-$. Hence, the $\Jcal$-hCS equations  (\ref{3.48}) on $(\Pcal^6, \Jcal )$ with
$\Fcal (V_3, \bar V_3)=0$ are equivalent to the SDYM equations on $\R^4$,
\begin{equation}\label{3.53}
F^-=0\quad\Leftrightarrow\quad\ve^{\al\be}F_{\al\da , \be\db}=0\ ,
\end{equation}
and any solution $A$ of the SDYM equations  (\ref{3.53}) defines a solution of the $\Jcal$-hCS equations  (\ref{3.48}) and vice versa.

\bigskip

\section{Twistor actions for Yang-Mills theory}

\noindent {\bf Graded twistor space $\Pcal^{6|2}$}. Recall that on  $\Pcal^{6}$ there are globally defined (3,0)-form $\Omega$ given by  (\ref{3.42}) and its complex conjugate  (0,3)-form $\bar\Omega$. Hence, the $\Jcal$-hCS action functional  (\ref{1.4}) is well defined on  $(\Pcal^6, \Jcal )$. However, if we substitute  (\ref{3.49}) into 
 (\ref{1.4}) then we obtain $S=0$ since (0,3)-part of Chern-Simons form CS($\Acal$) on $(\Pcal^6, \Jcal )$ vanishes if $\Acal_3=\bar\Acal_3=0$. To obtain a nontrivial Lagrangian, one can perform a gauge transformation, which will give some non-vanishing terms\footnote{Chern-Simons term CS($\Acal$) is not invariant under gauge transformations.} as it was done in~\cite{Costello, Skinner}. We will not follow this path here  because this way we can at best get the actions~\cite{Don}-\cite{Parkes} which have various limitations in comparison with the Chalmers-Siegel action~\cite{ChSi}.
 
 The action~\cite{ChSi} cannot be obtained without introducing additional degrees of freedom since it contains an extra propagating field $G_{\da\db}$. One of the possibilities for introducing additional fields is to consider vector bundles $\Ecal$ over $\Pcal^6$ that are not trivial after restriction to $\cp^1\hookrightarrow\Pcal^6$~\cite{Lei}. Another possibility is to consider a graded extension of the twistor space $(\Pcal^6, \Jcal )$ similar to the cases considered by Wolf~\cite{WolfPhD, Wolf2} and S\"amann~\cite{SaPhD, Saem} for the complex twistor space $\Pcal^3_\C$. We will use the second option and introduce a graded twistor space  $\Pcal^{6|2}$.
 
 The space $\Pcal^{6|2}$ is parametrized by bosonic coordinates on $\Pcal^{6}$ and by two anticommuting (fermionic) coordinates $\eta_i$,
 \begin{equation}\label{4.1}
 \eta_1\eta_2 + \eta_2\eta_1=0\ ,
\end{equation} 
generating the Grassmann algebra
\begin{equation}\label{4.2} 
\Lambda(\R^2)=\Lambda^0(\R^2)\oplus \Lambda^1(\R^2)\oplus\Lambda^2(\R^2)\ ,
\end{equation} 
where
\begin{equation}\label{4.3} 
{\bf 1}\cdot\R\in\Lambda^0(\R^2)\ ,\quad\eta_i\in\Lambda^1(\R^2)\ ,\quad i=1,2\und\eta:=\eta_1\eta_2\in\Lambda^2(\R^2)\ . 
\end{equation} 
In the algebra  (\ref{4.2})  one may introduce $\Z_2$-grading, 
\begin{equation}\label{4.4} 
\Lambda(\R^2)=\Lambda_0(\R^2)\oplus \Lambda_1(\R^2)\ ,
\end{equation} 
where
\begin{equation}\label{4.5} 
\Lambda_0(\R^2)=\Lambda^0(\R^2)\oplus \Lambda^2(\R^2)\and\Lambda_1(\R^2)=\Lambda^1(\R^2)\ .
\end{equation} 
We set $\gr f=0$ if $f\in\Lambda_0(\R^2)$ and $\gr f=1$ if $f\in\Lambda_1(\R^2)$, $\gr f$ is the Grassmann parity of $f$.

On the space $\Pcal^{6}$ we consider the space Gr$_{\Pcal^{6}}$ of locally defined functions (a sheaf) with values in the Grassmann algebra 
$\Lambda(\R^2)$. A manifold $\Pcal^{6}$ with the sheaf Gr$_{\Pcal^{6}}$ is a graded manifold $\Pcal^{6|2}= (\Pcal^{6},$ Gr$_{\Pcal^{6}}$)~\cite{Kos, Ber} that can be viewed as the trivial bundle $\Pcal^{6}\times\Lambda_1(\R^2)\to \Pcal^{6}$. Tangent spaces of $\Pcal^{6|2}$ are defined by the even vector fields  (\ref{3.33}),  (\ref{3.34}) together with the odd vector fields 
\begin{equation}\label{4.6} 
\pa^i:=\frac{\pa}{\pa\eta_i}\qquad\mbox{such that}\qquad\frac{\pa}{\pa\eta_1}\, \frac{\pa}{\pa\eta_2}+\frac{\pa}{\pa\eta_2}\, \frac{\pa}{\pa\eta_1}=0\ ,
\end{equation} 
commuting with the even vector fields on $\Pcal^{6}$. Respectively, the space of differential forms on $\Pcal^{6|2}$ has the local basis $\{E^a, \bar E^a, \diff\eta_i\}$ with commutation relations
\begin{equation}\label{4.7} 
\diff\eta_1\wedge\diff\eta_2=\diff\eta_2\wedge\diff\eta_1\ ,\quad E^a\wedge\diff\eta_i = \diff\eta_i \wedge E^a\und\bar E^a\wedge\diff\eta_i = \diff\eta_i \wedge \bar E^a
\end{equation} 
where $\{E^a, \bar E^a\}$ are given in (\ref{3.35}).

Recall that on $(\Pcal^{6}, \Jcal )$ there are globally defined forms $\Omega$ and $\bar\Omega$. Hence, on $\Pcal^{6|2}$ we can introduce a closed $(3|2)$-form
\begin{equation}\label{4.8}
\Imm\,\Omega\wedge\diff\eta_1\wedge\diff\eta_2
\end{equation} 
and the volume form 
\begin{equation}\label{4.9}
\frac{\im}{8}\,\Omega\wedge\bar\Omega\wedge\diff\eta =-\frac{\im}{2}\,\diff^4x\wedge\frac{\diff\la\wedge\diff\bar\la}{(1+\la\bar\la)^2}\wedge\diff\eta\ ,
\end{equation} 
where $\diff\eta =\diff\eta_1\wedge\diff\eta_2$.

\medskip

\noindent {\bf Chern-Simons type theory on $\Pcal^{6|2}$}. Let $\Ecal$ be a trivial rank $k$ complex vector bundle over $\Pcal^{6|2}$ and $\Acal$ a 
connection one-form on $\Ecal$. We choose the connection $\Acal$ depending on all coordinates on $\Pcal^{6|2}$ and having  no components along 
the Grassmann directions. The curvature $\Fcal$ of such $\Acal$ is
\begin{equation}\label{add1}
\Fcal = \Fcal^{\rm B} +\Fcal^{\rm F}=\diff^{\rm B}\Acal +\Acal\wedge\Acal+\diff^{\rm F}\Acal\ ,
\end{equation} 
where $\diff^{\rm B}$ is the bosonic part  (\ref{3.36}) of the exterior derivative $\diff =\diff^{\rm B}+\diff^{\rm F}$ and
\begin{equation}\label{add2}
\diff^{\rm F}=\diff\eta_i\partial^i\for\partial^i=\frac{\partial}{\partial\eta_i}
\end{equation} 
is the fermionic part of $\diff$.

Consider the action functional
\begin{equation}\label{4.10}
S=\int^{}_{\Pcal^{6|2}}\Imm\,\Omega\wedge\diff\eta\wedge\CS (\Acal )\ ,
\end{equation} 
where
\begin{equation}\label{4.11}
\CS (\Acal)=\tr (\Acal\wedge\diff^{\rm B}\Acal + \sfrac23\,\Acal\wedge\Acal\wedge\Acal )
\end{equation} 
is the Chern-Simons 3-form. Field equations following from   (\ref{4.10}) read
\begin{equation}\label{4.12}
\Imm\,\Omega\wedge\Fcal^{\rm B} =0\ ,
\end{equation} 
where $\Fcal^{\rm B}$ is defined in  (\ref{add1}).  From  (\ref{4.12}) it follows that
\begin{equation}\label{4.13}
\Re\,\Omega\wedge\Fcal^{\rm B} =0\ ,
\end{equation} 
since $\Omega$ is a (3,0)-form w.r.t. $\Jcal$,
\begin{equation}\label{4.14}
\Jcal\Omega = \im\,\Omega\quad\Leftrightarrow\quad    \Jcal\,\Imm\,\Omega =\Re\,\Omega\ .
\end{equation} 
Combining  (\ref{4.12}) and  (\ref{4.13}), we obtain
\begin{equation}\label{4.15}
\Omega\wedge\Fcal^{0,2}_{\rm B}=0\quad\Leftrightarrow\quad \Fcal^{0,2}_{\rm B}=0\ .
\end{equation} 
Note that from  (\ref{3.45}) and   (\ref{4.15}) it follows that~\cite{Har, Bunk}
\begin{equation}\nonumber
\Fcal^{\rm B} (V_1, \bar V_1) +\Fcal^{\rm B} (V_2, \bar V_2)=0\ .
\end{equation} 
The action functional  (\ref{4.10}) and solution to the equations  (\ref{4.12})-(\ref{4.15}) were considered in~\cite{Popov1, Har, Bunk}.

\medskip

\noindent {\bf Field equations on $\Pcal^{6|2}$}. Having given necessary ingredients, we may now consider $\Jcal$-hCS field equations (\ref{4.15}). These equations on the patch $\hat\Ucal_+=\Ucal_+\times\Lambda_1(\R^2)$ of $\Pcal^{6|2}$ read
\begin{equation}\label{4.16}
\bar V_\al\bar\Acal_\be - \bar V_\be\bar\Acal_\al + [\bar\Acal_\al , \bar\Acal_\be]=0\ ,\quad
\bar V_3\bar\Acal_\al - \bar V_\al\bar\Acal_3 + [\bar\Acal_3 , \bar\Acal_\al]- [\bar V_3 , \bar V_\al]\lrcorner\Acal =0\ ,
\end{equation}
where ``$\lrcorner$'' denotes the interior product of vector fields with differential forms. Here we used components of $\Acal$ in the expansion
\begin{equation}\label{4.17}
\Acal = \Acal_a E^a + \bar\Acal_a \bar E^a=\Acal_\al E^\al +\Acal_3 E^3 + \bar\Acal_\al \bar E^\al + \bar\Acal_3 \bar E^3\ .
\end{equation} 
As usual in the twistor approach, we work in a gauge in which $\bar\Acal_3\ne 0$ but the bosonic part of $\bar\Acal_3$ is zero.
Note that in general the gauge potential $\Acal$ in  (\ref{4.16}) and  (\ref{4.17}) can be expanded in the odd coordinates $\eta_i$ as
\begin{equation}\label{4.18}
\Acal = A + \eta_i\psi^i + \eta_1\eta_2\,G\ .
\end{equation}
For simplicity and more clarity we first consider the truncated case $\psi^i=0$ and discuss the case $\psi^i\ne0$ afterwards.

\noindent {\bf Remark.} The connection (\ref{4.18}) on the vector bundle $\Ecal$ over $\Pcal^{6|2}\cong \Pcal^{6}\times\Lambda_1(\R^2)$ takes
values in the Lie algebra $\mathfrak g$ of a semi-simple Lie group $\Gfr$. Note that maps from the space $\Lambda_1(\R^2)$ in (\ref{4.5}) to the group $\Gfr$ 
form a supergroup super-$T\Gfr$~\cite{Witten}, where $T\Gfr= \Gfr\ltimes\mathfrak g$ is the semi-direct product of $\Gfr$ and $\mathfrak g$. That is why the field 
$\Acal$ in (\ref{4.18}) can be considered as a connection on a super-$T\Gfr$ bundle $\Ecal'$ over the bosonic twistor space $\Pcal^{6}$. This kind of 
correspondence was found by Witten when studying Chern-Simons theories on 3-manifolds~\cite{Witten}.

From (\ref{3.33})-(\ref{3.35}) one concludes that components $\bar\Acal_\al$ and $\bar\Acal_3$ take values in the bundles $\Ocal (-1)$ and $\Ocal (2)$ over $\cp^1$ and $\Acal_\al$ and $\Acal_3$ take values in the complex conjugate bundles. This fixes the dependence of $\Acal$ on $\la$  and $\bar\la$ up to a gauge transformations (cf.~\cite{SaPhD, WolfPhD, Wolf1, Wolf2, Saem}). Namely, we obtain
\begin{equation}\nonumber
\Acal_\al{=}\ga\left\{\la^{\da} A_{\al\da}{+}\eta(\la^{\da} G_{\al\da}{+}\ga\la^{\da} \hat\la^{\db} \la^{\dot\ga} G_{\al\da\db\dot\ga})\right\}
{=}\ga\left\{\la^{\da} (A_{\al\da}{+}\eta B_{\al\da}){+}\eta\ga\la^{(\da} \hat\la^{\db} \la^{\dot\ga)}   G_{\al(\da\db\dot\ga)}\right\},
\end{equation}
\begin{equation}\nonumber
\bar\Acal_\al{=}\ga\,\jfr^\be_\al\,\left\{\hat\la^{\db} A_{\be\db}{+}\eta(\hat\la^{\db} G_{\be\db}{+}\ga\hat\la^{\db}\la^{\dot\ga}\hat\la^{\dot\sigma} 
G_{\be\db\dot\ga\dot\sigma})\right\}
{=}\ga\,\jfr^\be_\al\,\left\{\hat\la^{\db}(A_{\be\db}{+}\eta B_{\be\db}){+}\eta\ga\hat\la^{(\db}\la^{\dot\ga}\hat\la^{\dot\sigma)} 
G_{\be(\db\dot\ga\dot\sigma)})\right\},
\end{equation}
\begin{equation}\label{4.19}
\Acal_3=\eta\hat\la^{\db}\hat\la^{\dot\ga}G_{\db\dot\ga}\und\bar\Acal_3=-\eta\la^{\db}\la^{\dot\ga}G_{\db\dot\ga}\ ,
\end{equation}
where
\begin{equation}\label{4.20}
B_{\al\da}:=G_{\al\da}-\sfrac13\,\ve^{\db\dot\ga}( G_{\al\da\db\dot\ga}- G_{\al\db\dot\ga\da})
\end{equation}
and the coefficient fields $A_{\al\da}$, $G_{\al\da}$ and $G_{\al\da\db\dot\ga}$ do only depend on $x^{\al\da}\in\R^4$. Here $\la^{\da}, \hat\la^{\da}$ are given in 
(\ref{3.29}) and  (\ref{3.32}), $\eta = \eta_1\eta_2$, and parentheses denote normalized symmetrization with respect to the enclosed indices.

Substituting  (\ref{4.19}) into  (\ref{4.16}),  we obtain the equations 
\begin{equation}\label{4.21}
 G_{\al(\da\db\dot\ga)}=\nabla_{\al(\da}G_{\db\dot\ga)}\ ,
\end{equation}
showing that $G_{\al(\da\db\dot\ga)}$ are composite fields describing no independent degrees of freedom. Other nontrivial equations following from (\ref{4.16})  after substituting  (\ref{4.19})   read
\begin{equation}\label{4.22}
\ve^{\al\be}\left[\pa_{\al\da} + A_{\al\da}, \pa_{\be\db} + A_{\be\db}\right]=\ve^{\al\be}F_{\al\da , \be\db}=0\ ,
\end{equation}
\begin{equation}\label{4.23}
\ve^{\al\be}\na_{\al\da} B_{\be\db}=0\ ,
\end{equation}
\begin{equation}\label{4.24}
\ve^{\da\db}\na_{\al\da} G_{\db\dot\ga}=0\ .
\end{equation}
We see that (\ref{4.22}) coincide with the SDYM equations on $\R^4$ and (\ref{4.23}) are the linearized SDYM equations for
\begin{equation}\label{4.25}
\de A_{\be\db}= B_{\be\db}\ .
\end{equation}
Hence, $B_{\al\da}$ is a tangent vector at $A_{\al\da}$ to the solution space of the SDYM equations. It is a secondary field (a symmetry) depending on $A_{\al\da}$ and for simplicity we neglect it by choosing $B_{\al\da}=0$. The rest equations  (\ref{4.22}) and  (\ref{4.24}) are the Chalmers-Siegel equations describing the self-dual gauge potential $A_{\al\da}$ and the anti-self-dual field $G_{\al\da , \be\db}=\ve_{\al\be}G_{\da\db}$ propagating in the self-dual background.

The action functional associated with  (\ref{4.22}) and  (\ref{4.24}) is given by
\begin{equation}\label{4.26}
S_{\mathrm{sd}}=2\int_{\R^4}\diff^4x\ \tr (G^{\da\db}f_{\da\db})
\end{equation}
with $f_{\da\db}$ given by  (\ref{3.52}). This action can be obtained from  (\ref{4.10}) after splitting,
\begin{equation}\label{4.27}
\Acal = X + \eta Y\ ,
\end{equation}
into ordinary bosonic and even nilpotent parts, using the formula\footnote{Recall that  in all formulae here $\diff^{\rm B}$ is the ordinary bosonic exterior derivative.}
\begin{equation}\label{4.28}
\CS (X+\eta Y)=\CS (X) + 2\eta\,\tr (Y\wedge\Fcal (X)) - \eta\,\diff^{\rm B} (\tr (X\wedge Y))
\end{equation}
and integrating over the nilpotent coordinate $\eta$ and over $\cp^1\hookrightarrow\Pcal^{6|2}$.

\medskip

\noindent {\bf Extra terms.} As we mentioned earlier, the general expansion (\ref{4.18}) of connection $\Acal$ in odd coordinates $\eta_i$ contains fermionic fields $\psi^i (x^{\al\da})$ which we consider now. Expansion (\ref{4.18}) can be written in components as
\begin{equation}\label{4.29}
\Acal_\al = \ga\, \la^{\da} A_{\al\da}(\eta_1, \eta_2)\and \Acal_3 = \hat\la^{\da}\hat \la^{\db} G_{\da\db}(\eta_1, \eta_2)\ ,
\end{equation}
where
\begin{equation}\label{4.30}
A_{\al\da}(\eta_1, \eta_2)=A_{\al\da}+\eta_i(\psi^i_{\al\da} + \ga\hat\la^{\db} \la^{\dot\ga}\psi^i_{\al(\da\db\dot\ga )}) + 
\eta_1\eta_2(B_{\al\da} + \ga\hat\la^{\db} \la^{\dot\ga}G_{\al(\da\db\dot\ga )}) \ ,
\end{equation}
\begin{equation}\label{4.31}
G_{\da\db}(\eta_1, \eta_2)=\eta_i \psi^i_{\da\db} +\eta_1\eta_2 G_{\da\db}  \ .
\end{equation}
For $\bar\Acal_\al$ and $\bar\Acal_3$ we have
\begin{equation}\label{4.32}
\bar\Acal_\al = \ga\,\jfr^\be_\al\,\hat\la^{\db}A_{\be\db}(\eta_1, \eta_2)\and \bar\Acal_3 = -\la^{\da}\,\la^{\db}\,G_{\da\db}(\eta_1, \eta_2)\ .
\end{equation}
Substituting  (\ref{4.29})-(\ref{4.32})  into the equations  (\ref{4.16}), we obtain the equations  (\ref{4.21})-(\ref{4.25}) and additional equations
\begin{equation}\label{4.33}
\ve^{\al\be}\nabla_{\al\da}\psi^1_{\be\db}=0\and \psi^2_{\be\db}=0\ ,
\end{equation}
\begin{equation}\label{4.34}
\ve^{\da\db}\nabla_{\al\da}\psi^1_{\db\dot\ga}=0\and \psi^1_{\al(\da\db\dot\ga)}= \nabla_{\al(\da}\psi^1_{\db\dot\ga)}\ ,
\end{equation}
\begin{equation}\label{4.35}
\psi^2_{\db\dot\ga}=0\and \psi^2_{\al(\da\db\dot\ga)}=\nabla_{\al(\da}\psi^2_{\db\dot\ga)}=0\ .
\end{equation}
From  (\ref{4.23}) and  (\ref{4.33}) we see that $B_{\al\da}$ and $\psi^1_{\al\da}$ are even and odd solutions of the linearized SDYM equations and $\psi^1_{\db\dot\ga}$ in 
 (\ref{4.34}) is an odd solution to the linearized form of equation  (\ref{4.24}) for $\de G_{\db\dot\ga}$. Thus, the general form (\ref{4.18}) of $\Acal$ reduces the $\Jcal$-hCS equations (\ref{4.16}) to the Chalmers-Siegel equations (\ref{4.22}) and (\ref{4.24}) together with their linearized form, solutions of which describe even and odd tangent vectors to the solution space.
 
 \medskip

\noindent {\bf Full Yang-Mills.} So far, we have shown that the Chalmers-Siegel action  (\ref{4.26}) for SDYM theory can be obtained from the Chern-Simons type action (\ref{4.10}) on the graded twistor space $\Pcal^{6|2}$. It is known that the action  (\ref{4.26}) is a limit of the full Yang-Mills action for small coupling constant $g^{}_{\rm YM}$. Namely, let us modify the action  (\ref{4.26}) by adding the term
\begin{equation}\label{4.36}
S_\ve = -\ve^2\,\int_{\R^4}\diff^4x\,\tr (G^{\da\db}G_{\da\db})\ ,
\end{equation}
so that
\begin{equation}\label{4.37}
S_{\rm{tot}} = S_{\rm{sd}}+S_\ve =  2\int_{\R^4}\diff^4x\,\tr (G^{\da\db}f_{\da\db} -\sfrac12\, \ve^2\, G^{\da\db}G_{\da\db})    \ .
\end{equation}
Here $\ve$ is some small parameter. Variation of $S_{\rm{tot}}$ w.r.t. $G_{\da\db}$ gives
\begin{equation}\label{4.38}
G_{\da\db}= \frac{1}{\ve^2}\, f_{\da\db}\ .
\end{equation}
Substituting  (\ref{4.38}) back into  (\ref{4.37}), we obtain 
\begin{equation}\label{4.39}
\begin{aligned}{}
S_{\rm{tot}} =& \frac{1}{\ve^2}\, \int_{\R^4}\diff^4x\,\tr (f_{\da\db}f^{\da\db})=  \frac{1}{2\ve^2}\,  \int_{\R^4}\tr (F^-\wedge F^-)\\
=&-\frac{1}{4\ve^2}\,  \int_{\R^4}\tr (F\wedge \ast F) +\frac{1}{4\ve^2}\,  \int_{\R^4}\tr (F\wedge F)\ .
\end{aligned}
\end{equation}
Hence, the action  (\ref{4.39}) is equivalent to the standard Yang-Mills action
\begin{equation}\label{4.40}
S_{\rm{YM}} =- \frac{1}{4g^2_{\rm{YM}}}\, \int_{\R^4}\tr (F\wedge \ast F) \ ,
\end{equation}
with the coupling constant $g_{\rm{YM}} =\ve$, plus the topological term. Therefore, for obtaining the Yang-Mills action (\ref{4.40}) we should derive the term (\ref{4.36}) from the twistor space. 

\medskip

\noindent {\bf Twistor action for full Yang-Mills.} Recall that $\eta = \eta_1\eta_2$, where $\eta_1$ and $\eta_2$ are real Grassmann variables. Consider a connection $\Acal$ depending on $\eta$ as written in (\ref{4.19}).\footnote{We do not consider the more general dependence  (\ref{4.29})-(\ref{4.32}) since we only want to show that one {\it can} obtain the action (\ref{4.40}) from the twistor space. Consideration of (\ref{4.29})-(\ref{4.32}) will give the Yang-Mills theory with its infinitesimal symmetries as we saw in the case of the SDYM equations.} It does not have components along the Grassmann directions but the mixed components of the curvature,
\begin{equation}\label{add3}
\Fcal ^{\rm F}= \diff^{\rm F}\Acal= (\partial^i\Acal_a)\diff\eta_i\wedge E^a + (\partial^i\bar\Acal_a)\diff\eta_i\wedge \bar E^a  = \Fcal^i_a\diff\eta_i\wedge E^a + 
\bar\Fcal^i_a\diff\eta_i\wedge \bar E^a    \ ,
\end{equation} 
do not vanish. In particular, for restriction of $\Fcal^{\rm F}$ to $\cp^{1|2}\hookrightarrow\Pcal^{6|2}$ we have
\begin{equation}\label{add4}
\Fcal ^{\rm F}_{|\cp^{1|2}}=\Fcal^i_\la\,\diff\eta_i\wedge\diff\la +\Fcal^i_{\bar\la}\,\diff\eta_i\wedge\diff{\bar\la }\ ,
\end{equation} 
where 
\begin{equation}\label{4.41}
\Fcal^i_\la =-\ve^{ij}\eta_j\ga^2\hat\la^{\da}\hat\la^{\db}\, G_{\da\db}\and\Fcal^i_{\bar\la} =\ve^{ij}\eta_j\ga^2\la^{\da}\la^{\db}\, G_{\da\db}\ .
\end{equation}
Using (\ref{4.41}), we can introduce the gauge invariant functional
\begin{equation}\label{4.43}
\frac{\im\ve^2}{8}\, \int_{\Pcal^{6|2}}\Omega\wedge \bar\Omega\wedge\diff\eta_1\wedge\diff\eta_2\,\ve_{ij}\, g^{\la\bar\la}\tr (\Fcal^i_\la\Fcal^j_{\bar\la})\ ,
\end{equation}
where
\begin{equation}\label{4.44}
\diff s^2_{\cp^1}=g_{\la\bar\la}E^\la\otimes E^{\bar\la}\quad\Leftrightarrow\quad g_{\la\bar\la}=\ga^2\und g^{\la\bar\la}=\ga^{-2}\ .
\end{equation}
Integrating $\tr (\Fcal^i_\la\Fcal^j_{\bar\la})$ in (\ref{4.43}) over fermionic coordinates and over $\cp^1\hookrightarrow\Pcal^{6|2}$, we obtain the functional $S_\ve$ given by  (\ref{4.36}). Hence, adding the {\it local} term given by (\ref{4.43}) to the Chern-Simons type Lagrangian in (\ref{4.10}), we obtain the full Yang-Mills action (\ref{4.40}).

\section{Conclusions}

In this paper we considered graded twistor space $\Pcal^{6|2}$ with a non-integrable almost complex structure $\Jcal$ and $\Jcal$-holomorphic Chern-Simons theory on $\Pcal^{6|2}$. It was shown that under some assumptions this theory is equivalent to self-dual Yang-Mills theory on $\R^4$. In our discussion we tried to be close to the consideration of the papers~\cite{Wolf2, Saem}, where $\Ncal <4$ SDYM theories were derived from holomorphic Chern-Simons theory on complex supertwistor spaces. We have also shown that the full Yang-Mills action in $\R^4$ can be obtained from a twistor action on $\Pcal^{6|2}$ with a locally defined Lagrangian. We did not pursue the goal of studying all these tasks in full generality. We wanted to show the principal possibility of obtaining actions for Yang-Mills and its self-dual subsector from a twistor action. Examining all aspects of the model requires additional efforts.

\bigskip
\noindent {\bf Acknowledgements}

\noindent
This work was supported by the Deutsche Forschungsgemeinschaft grant LE~838/19.

\end{document}